\documentclass{PoS}

\newcommand{\hi}{{\sc H\,i}}

\title{Panoramic Radio Astronomy: Wide-field 1-2 GHz research on galaxy evolution}

\ShortTitle{PRA2009}

\author{George Heald \& Paolo Serra\\
        The Netherlands Foundation for Radio Astronomy (ASTRON)\\
        E-mail: \email{heald@astron.nl}}

\abstract{In this contribution we give a brief overview of the Panoramic Radio Astronomy (PRA) conference held on 2-5 June 2009 in Groningen, the Netherlands. The conference was motivated by the on-going development of a large number of new radio telescopes and instruments which, within a few years, will bring a major improvement over current facilities. Interferometers such as the Expanded Very Large Array (EVLA), Australian SKA Pathfinder (ASKAP), Allen Telescope Array (ATA), Karoo Array Telescope (MeerKAT) and the Aperture Tile In Focus (APERTIF) upgrade to the Westerbork Synthesis Radio Telescope (WSRT) will provide a combination of larger field of view and increased simultaneous bandwidth, while maintaining good collecting area and angular resolution. They will achieve a survey speed 10-50 times larger at 1-2 GHz than the current possibilities, allowing for the first time optical-like all-sky extra-galactic surveys at these frequencies.

Significant progress will be made in many fields of radio astronomy. In this conference we focused on research into the evolution of galaxies over the past few Gyr. In particular, wide-field observations at 1-2 GHz will provide an unprecedented panoramic view of the gas properties and star formation in galaxies, embedded in their environment, from $z\sim$0.2-0.5 to the present. Within the framework of our current knowledge of the galaxy population at $z<0.5$, we discussed: the key science questions that the new telescopes will permit us to answer in combination with complimentary work at other wavelengths; the observing modes and analysis strategies which will allow us to most efficiently exploit the data; and the techniques for most effectively coping with the huge volume of survey products, so far unusual for the radio community. Emphasis was placed on the complementarity of the upcoming facilities and on their role in paving the way for the technological development and science goals of the Square Kilometre Array.}

\FullConference{Panoramic Radio Astronomy: Wide-field 1-2 GHz research on galaxy evolution - PRA2009\\
		 June 02 - 05 2009\\
		 Groningen, the Netherlands}

\begin{document}

\section{Motivation}

In the coming few years, a series of new radio telescope facilities will be built or significantly upgraded. These telescopes become feasible mainly thanks to advances both in receiver technology and, perhaps most importantly, in computing. As can be seen throughout these proceedings, the strength of the new observatories when compared to their predecessors is the ability to perform deep, wide-field surveys of the whole sky, truly enabling the new era of {\it Panoramic Radio Astronomy}.

The Panoramic Radio Astronomy (PRA) conference was held in Groningen (the Netherlands) on 2-5 June 2009. We focused in particular on the following upcoming facilities: Aperture Tile In Focus [APERTIF; a significant upgrade to the Westerbork Synthesis Radio Telescope (WSRT)] \cite{oosterloo_tv}, the Australian SKA Pathfinder (ASKAP) \cite{braun_tv}, the Allen Telescope Array (ATA) \cite{leeuwen_tv}, the Expanded Very Large Array (EVLA) \cite{rupen_tv}, and the Karoo Array Telescope (MeerKAT) \cite{jonas_tv}. Although these will all individually be powerful instruments, with significant impact on our understanding of the Universe, the scientific plans are still somewhat in their formative phases, some more advanced than others (and all have progressed even further in the time since the PRA conference). When organizing this meeting, it was our feeling that little thought had been given so far to the combined strength of the next generation of radio telescopes, and we sought to bring together the experts to advance that conversation.

\section{The PRA conference}

The overarching scientific goal of the PRA conference was to address the scientific application of these new instruments to questions regarding the evolution of galaxies (in particular the aspects related to their gas content and star formation) over the past few billion years. More specifically, we listed the following topics as central to the conference:
\begin{itemize}
\item Scope, depth and design of \hi\ wide area surveys
\item Evolution of the \hi\ mass function and its dependence on morphological type and environment
\item Evolution of galaxy scaling relations out to $z\sim0.2$
\item The evolution of star formation and its relation to gas content in galaxies
\item Wide field-of-view, deep \hi\ observations of individual fields - nearby clusters, groups and galaxies
\item Continuum surveys: star-forming-galaxies and the role of AGN activity
\item Polarisation and magnetic fields in nearby galaxies
\end{itemize}

We drew together experts on the telescope facilities themselves, researchers into the observational and theoretical aspects of the scientific questions to be addressed, and astronomers with multi-wavelength experience who could provide perspective into how the future large radio surveys will complement existing and future surveys at other frequencies. It was gratifying that we had plenty of contributions on the expected topics, but even more so that we also had many contributions dealing with a larger variety of projects -- implying that the community is waiting for these telescopes to provide data for a myriad of different experiments!

\begin{figure}
\centering
\includegraphics[width=\textwidth]{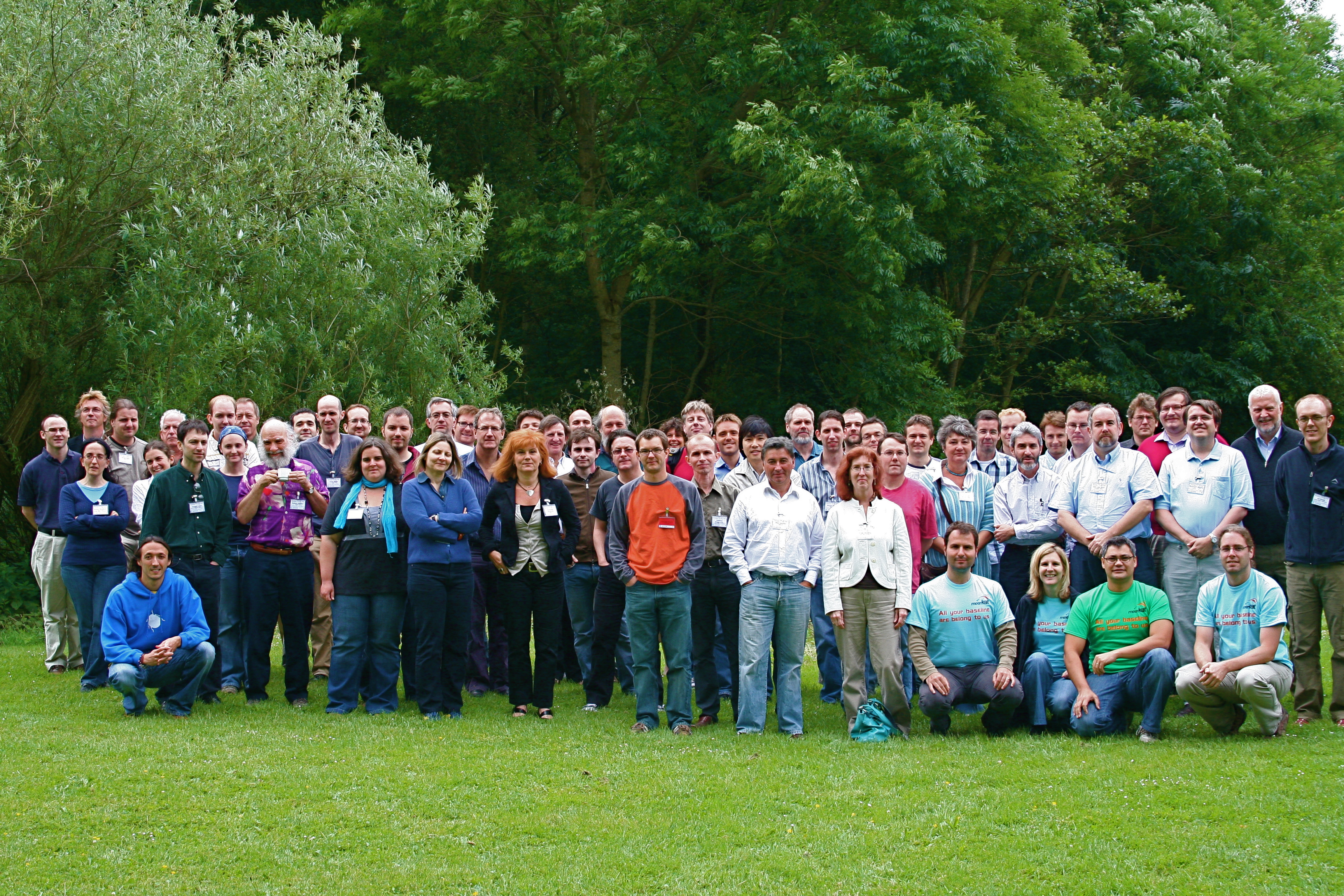}
\caption{Participants of the Panoramic Radio Astronomy conference, enjoying a beautiful spring day in Groningen.}
\label{fig:pic}
\end{figure}

In addition to the standard conference setup, featuring a program of talks and a poster session, we incorporated two organised discussion sessions at the beginning and at the end of the conference. These sessions were chaired by Thijs van der Hulst (to whom we are grateful for his deft guidance!), and featured a panel made up of five ``representatives'' from the upcoming telescopes listed in \S\,1.

\section{Outcomes}

During the conference, it quickly became clear that in the next decade there will be a remarkable similarity in observational capabilities in the Northern and Southern hemispheres. Each hemisphere will host at least one radio facility able to perform relatively rapid, moderately deep all-sky surveys in the 1-2 GHz regime; and one radio facility able to perform deep, but narrower-field observations at the same (and higher) frequencies. In the North, a highly complementary pair is Apertif (a focal plane array upgrade to the Westerbork Synthesis Radio Telescope) and the EVLA (a significant upgrade to the Very Large Array); in the South we will have ASKAP (a brand-new focal plane array-based SKA precursor) and MeerKAT (a brand-new single-receiver based SKA precursor). The Allen Telescope Array, also in the northern hemisphere, brings together the benefits of wide-field imaging and flexibility in the observing frequency. The two ``pairs'' are extremely similar in terms of their angular resolution, sensitivity, and field of view (see their respective contributions), making truly {\it all-sky surveys} and the crucial narrower deep surveys (and extension to higher frequencies) all accessible in the radio. The types of scientific programs that can be carried out, and were discussed at length during the conference, include:
\begin{itemize}
\item large and deep \hi\ and OH surveys to redshifts of $z\sim1$ and 1.6, respectively
\item determination of the variation of the \hi\ mass function, especially at the faint end, with redshift
\item a detailed picture of the faint radio continuum source population
\item studies of the gas content in galaxies and their environment at redshifts up to about $z\sim0.1$
\item clarification of the magnetic field (and cosmic ray) properties in galaxies, both nearby and at cosmological distances
\end{itemize}
We refer to the individual contributions in these proceedings for excellent descriptions of these, and other, scientific applications.

In the slightly more distant future, the radio community is looking forward to, and many are working diligently toward, the full Square Kilometre Array (SKA), which is to be sited in either Australia or South Africa. Several talks during the conference focused on how the programs carried out by the next generation of telescopes (the Pathfinders and Precursors) will further motivate and, we strongly hope, enhance the SKA science case. We were happy that Joe Lazio joined us at the conference and presented the highlights of the scientific mission of the SKA \cite{lazio_tv}.

As mentioned earlier, the conference was bookended by two discussion sessions, which first motivated, and later summarized, the science questions addressed throughout the conference. In particular the second discussion session was quite lively, and it became clear that a more pro-actively cooperative participation (both between and within the two Hemispheres) has been motivated by the conference. This is true not only from the scientific point of view, but also regarding practical issues such as data format/products in order to maximise the exploitation of the new facilities and widen the radio astronomical community.

\bigskip

Finally, we could not avoid organising an excursion to Exloo, the location of the LOFAR core. At the time of the conference, the first full station was in place and the ``superterp'' -- the heart of LOFAR's core -- was nearly ready for antennas to be placed on it. We also visited the facility where the highband (120--240\,MHz) antennas are being assembled. While LOFAR is certainly outside of the nominal frequency range covered by the PRA conference, it is also an SKA pathfinder, and has strong synergy with the higher frequency telescopes \cite{morganti_tv}. Since the conference, the buildup of LOFAR has continued apace, but that's a story for another time\ldots

\section{Participants}

The conference participants (in total, 79) came from a large range of countries, as expected based on the geographical distribution of the telescopes under discussion and indeed the global nature of the SKA project, a longer-term goal of the community that drives much of the current development. 
More than half of the participants were students and young researchers, reflecting the fact that the community is being reborn into a new era!

\acknowledgments{The organizers gratefully acknowledge the generous financial support of ASTRON, CSIRO-ATNF, the University of Cape Town, and RadioNet. We thank Lucas Elting for his invaluable help with the computers and projection equipment; the Hampshire hotel for their hospitality; Henk Paarhuis, Ger de Bruyn and Michiel Brentjens for their enthusiasm in guiding us through the LOFAR project; and especially Marjan Tibbe for her tireless efforts in organizing all of the details, large and small. Finally, our thanks to all of the participants of the meeting for making this a very enjoyable and motivating conference!}


\begin{thebibliography}{99}
\bibitem{braun_tv} 
R.~Braun, \emph{Panoramic Surveys of the Radio Sky with the Australian SKA Pathfinder}, in proceedings of \emph{PRA2009}, \pos{PoS(PRA2009)002}.

\bibitem{jonas_tv}
J.~Jonas, \emph{The MeerKAT SKA precursor telescope}, in proceedings of \emph{PRA2009}, \pos{PoS(PRA2009)004}.

\bibitem{leeuwen_tv}
D.~Backer et al., \emph{The Allen Telescope Array: The First Widefield, Panchromatic, Snapshot Radio Camera}, in proceedings of \emph{PRA2009}, \pos{PoS(PRA2009)005}.

\bibitem{oosterloo_tv}
T.~Oosterloo, \emph{The latest on Apertif}, in proceedings of \emph{PRA2009}, \pos{PoS(PRA2009)006}.

\bibitem{rupen_tv}
M.~Rupen, \emph{The EVLA: Progress and Prospects}, in proceedings of \emph{PRA2009}, \pos{PoS(PRA2009)003}.

\bibitem{lazio_tv}
J.~Lazio, \emph{The Square Kilometre Array}, in proceedings of \emph{PRA2009}, \pos{PoS(PRA2009)058}.

\bibitem{morganti_tv}
R.~Morganti, \emph{Continuum Surveys with LOFAR}, in proceedings of \emph{PRA2009}, \pos{PoS(PRA2009)040}.
\end{thebibliography}
\end{document}